\documentclass{aa}

\usepackage{txfonts}
\usepackage{graphicx}

\usepackage{mathptmx}
\usepackage[english]{babel}
\usepackage{amsmath}

\usepackage{natbib,twoopt}
\bibpunct{(}{)}{;}{a}{}{,} 
\makeatletter
\newcommandtwoopt{\citeads}[3][][]{\href{http://adsabs.harvard.edu/abs/#3}%
{\def\hyper@linkstart##1##2{}%
\let\hyper@linkend\@empty\citealp[#1][#2]{#3}}}
\newcommandtwoopt{\citepads}[3][][]{\href{http://adsabs.harvard.edu/abs/#3}%
{\def\hyper@linkstart##1##2{}%
\let\hyper@linkend\@empty\citep[#1][#2]{#3}}}
\newcommandtwoopt{\citetads}[3][][]{\href{http://adsabs.harvard.edu/abs/#3}%
{\def\hyper@linkstart##1##2{}%
\let\hyper@linkend\@empty\citet[#1][#2]{#3}}}
\newcommandtwoopt{\citeyearads}[3][][]%
{\href{http://adsabs.harvard.edu/abs/#3}
{\def\hyper@linkstart##1##2{}%
\let\hyper@linkend\@empty\citeyear[#1][#2]{#3}}}
\makeatother

\newcommand{\be}{\begin{equation}}
\newcommand{\ee}{\end{equation}}
\newcommand{\bea}{\begin{eqnarray}}
\newcommand{\eea}{\end{eqnarray}}
\newcommand{\bi}{\begin{itemize}}
\newcommand{\ei}{\end{itemize}}
\renewcommand{\d}{{\rm d}}
\renewcommand{\v}[1]{\mathbf{#1}}
\renewcommand{\i}{\item}
\newcommand{\erf}{{\rm erf}}
\newcommand{\ssymb}{Section~}
\newcommand{\appsymb}{Appendix~}
\newcommand{\figsymb}{Figure~}
\newcommand{\tabsymb}{Table~}
\newcommand{\figref}[1]{\figsymb\ref{#1}}
\newcommand{\tabref}[1]{\tabsymb\ref{#1}}
\renewcommand{\sec}[1]{\ssymb\ref{#1}}             
\newcommand{\seca}[1]{{\appsymb~\ref{#1}}}
\newcommand{\rout}{r_{\rm out}}
\newcommand{\invlam}{\kappa}

\begin{document}

\title{MODA: a new algorithm to compute optical depths in 
       multi-dimensional hydrodynamic simulations}

\author{Albino Perego\inst{1,3}, 
        Emanuel Gafton\inst{2}, 
        Rub{\'{e}}n Cabez{\'{o}}n\inst{3},
        Stephan Rosswog\inst{2} \and
        Matthias Liebend{\"{o}}rfer\inst{3}}

\institute{TU Darmstadt, Institut f{\"{u}}r Kernphysik 
           Theoriezentrum, Schlossgartenstr. 2, D-64289 Darmstadt, Germany\\
           \email{albino.perego@physik.tu-darmstadt.de}
           \and
           The Oskar Klein Centre, Department of Astronomy, 
           AlbaNova, Stockholm University, SE-106 91 Stockholm, Sweden
           \and
           Department of Physics, University of Basel, 
           Klingelbergst. 82, CH-4056 Basel, Switzerland
           }
        
\date{Received ,; accepted , }
        
\titlerunning{MODA}
\authorrunning{Perego et al.}

\abstract
{}    
{We introduce a new algorithm for the calculation of multidimensional 
optical depths in approximate radiative transport schemes, 
equally applicable to neutrinos and photons.
Motivated by (but not limited to) neutrino transport in three-dimensional 
simulations of core-collapse supernovae and neutron star mergers, 
our method makes no assumptions about the geometry of the matter
distribution, apart from expecting optically transparent boundaries.}
{Based on local information about opacities, the algorithm 
figures out an escape route that tends to minimize the optical depth
without assuming any pre-defined paths for radiation.
Its adaptivity makes it suitable for a variety of 
astrophysical settings with complicated geometry (e.g., core-collapse supernovae, 
compact binary mergers, tidal disruptions, star formation, etc.).
We implement the MODA algorithm into both a Eulerian hydrodynamics code with
a fixed, uniform grid and into an SPH code where we make use a tree structure that
is otherwise used for searching neighbours and calculating gravity.}
{In a series of numerical experiments, we compare the MODA results with 
analytically known solutions. We also use snapshots from actual 3D simulations
and compare the results of MODA with those obtained with other methods such as the
global and local ray-by-ray method. It turns out that MODA achieves 
excellent accuracy at a moderate computational cost.
In an appendix we also discuss implementation details 
and parallelization strategies.} 
{}

\keywords{methods: numerical -- neutrinos -- radiative transfer}

\maketitle          
%
%
\section{Introduction}
\label{sec:intro}

Recent years have seen an enormous increase 
in the physical complexity of multi-dimensional, 
astrophysical simulations. 
This is particularly true for three-dimensional
radiation-hydrodynamic simulations as exemplified 
by recent advances in, say, star formation 
\citep[e.g.,][]{bate12}
and core-collapse supernovae 
\citep[e.g.,][and references therein]{Janka2012,Burrows2013,ott13}.
In the context of radiative transfer problems, the optical depth ($\tau$) is of central
importance. When it is calculated from the radiation production site up to the
transparent edges of the system, it counts the average number of interactions
experienced by a radiation particle before it can finally escape. 
The optical depth allows the distinction between
diffusive ($\tau \gg 1$), semi-transparent ($\tau \sim 1$), and transparent
($\tau \ll 1$) regimes. The surface where $\tau = 2/3$ represents the last
interaction surface and is often called \textit{neutrinosphere} in the 
case of neutrino radiation (or \textit{photosphere} in the case of photons).

Codes that solve the Boltzmann equations, through either finite difference
(e.g., the discrete ordinate method, see \citealt{liebendoerfer04} in 1D, 
\citealt{ott08} in 2D),
or spectral methods (e.g., the spherical harmonic method, see \citealt{peres13b}),
Monte-Carlo codes \citep[e.g.,][]{abdikamalov12}, and
most approximative schemes such as flux-limited diffusion 
\citep[e.g.,][]{whitehouse05,swesty09} or M1 schemes 
\citep[e.g.,][]{shibata2011,OConnor2013},
do not need to compute $\tau$ separately, 
since it is a physical quantity that results from the algorithm itself.
Other codes, such as that of \cite{kuroda12}, which also employs the 
M1 closure of the transport equations using a variable Eddington factor,
calculate $\tau$ assuming that radiation moves along radial paths. 

Some approximated radiation transport schemes, 
like the Isotropic Diffusion Source Approximation \citep{Liebendorfer2009},
require the calculation of $\tau$ to determine the location of the 
neutrinospheres. Also, light-bulb methods \citep[e.g.][]{Nordhaus2010,Hanke2012,Couch2013a} 
and ray-tracing methods \citep[e.g.,][]{kotake03,Caballero2009,Surman2013}, 
in which the neutrino flux intensity in the free streaming region 
is computed from the inner boundary luminosities or by assuming black body emission at
the neutrino decoupling surface, often demand the computation of $\tau$, 
at least where $\tau \lesssim 1$.

Codes that treat radiation with leakage schemes 
\citep[going back to][]{vanriper81,bludman1982,cooperstein1986} 
need to calculate the optical depth explicitly everywhere inside the computational domain,
since it is directly related to the diffusion time scale.
In grid-based codes, this is traditionally achieved using either
a global or a local ray-by-ray (RbR) approach, while in meshless codes
like SPH one often interpolates the mean free path to a grid, 
calculates $\tau$ just like in the grid-based approach, and then interpolates
it back to the SPH particles.
The global RbR method employs a number of pre-defined radial rays,
centered in one element of the domain, to calculate the optical depth, and is most effective 
in geometries with spherical symmetry, such as core-collapse supernovae
\citep[CCSNe;][]{peres13,ott13} and isolated (relativistic) stars \citep{galeazzi13}.
The local RbR method, on the other hand, integrates the optical depth equation along 
pre-defined rays starting at each point of the grid. This approach is computationally
more expensive, but it is able
to handle more complex geometries, and has been used for instance in
Newtonian neutron star mergers \citep{ruffert96,rosswog03} and
neutron star--black hole mergers \citep{deaton13}.
More recently, neutrino leakage schemes have been adapted to the formalism of
general relativity \citep{sekiguchi10,galeazzi13} and 
used in simulations of neutron star mergers \citep{sekiguchi11}.

Our work is motivated by questions related to core-collapse supernovae and neutron star mergers,
but our algorithm makes no assumption about the geometry, radiation path or the astrophysical
scenario, and could therefore be readily used in other contexts (e.g., photons in stellar atmospheres).\\

\noindent This paper is organised as follows.
\sec{sec:method} is devoted to a brief review of the
concept of optical depth and an outline of our main hypotheses
(\sec{sec:method definitions assumptions}), together with 
a general description of our new algorithm (\sec{sec:method algorithm}).
In \sec{sec:implem} we describe its implementation in grid-based 
(\sec{sec:implem grid}) and SPH schemes
(\sec{sec:implem SPH}).
\sec{sec:tests analytic} presents a number of tests with analytically known solutions, while 
\sec{sec:tests astroph} compares the results of various numerical methods applied 
to snapshots from grid-based (\sec{sec:tests astroph grid}) and 
SPH (\sec{sec:tests astroph SPH}) simulations.
Performance and scaling of MODA are discussed in \sec{sec:performance}.
Finally, our results are summarized in \sec{sec:discussion}, while
technical details about the implementation and the parallelization are 
presented in \seca{sec:app technical}, 
and in \seca{sec:app parallelization}.

%
%
\section{Method}
\label{sec:method}

\subsection{Definitions and assumptions}
\label{sec:method definitions assumptions}

\noindent The optical depth is a quantitative measure of the interaction 
between radiation and matter between two points, related on infinitesimal scales 
to the mean free path $\lambda$. Along a path $\gamma$ connecting two points 
$A$ and $B$, the optical depth is defined as 
\be\label{eq:general optical depth}
\tau_\gamma(A \rightarrow B) = \int_{\gamma:\/A \rightarrow B} \frac{\d s}{\lambda(s)},
\ee 
\noindent where $\lambda(s)$ is the local mean free path, and $\d s$ is an infinitesimal displacement 
along the chosen path, $\gamma$. Eq.~\eqref{eq:general optical depth} already contains the physical
interpretation of $\tau$: being related to the inverse mean free path, it counts the average
number of interactions between radiation and matter along the path $\gamma$.

Even though each radiation particle travels and interacts in its own way and along its own path,
for radiation transport problems involving astrophysical (i.e. macroscopic) objects, we are
interested in a \textit{statistical} description of the radiation and in its \textit{global} behaviour 
(i.e., from the site of production up to the point where radiation can escape).
We first notice that from a microscopic point of view\footnote{By 
macro-/microscopic we mean large/small in comparison to the local mean free path $\lambda$.}, 
the production occurs isotropically, i.e., radiation can be emitted with equal probabilities in 
any direction. On the macroscopic scale, however, the properties of matter influence 
the behaviour of radiation and break the isotropy of the emission: 
if a radiation particle is emitted towards a region of decreasing mean free path, 
it will likely interact again with matter, changing its original propagation direction; 
on the other hand, if it is emitted towards a direction of increasing mean free path, 
it will probably move away freely from the production site. 
This simple consideration suggests that, \textit{even if radiation is emitted 
locally in an isotropic way, macroscopically it moves preferentially 
towards regions of larger mean free path}.
Second, \textit{the path followed by radiation particles 
between two points is not therefore necessarily straight}: if the two points are 
separated by a relatively opaque region, radiation emitted in that
direction will be likely scattered or absorbed 
(and, eventually, re-emitted in another direction). 
Thus, the global, statistical behaviour of the radiation moving between those two points 
will be to bypass that opaque region
along a non straight path, characterized by larger mean free paths.

Ultimately, when radiation particles have reached locations where the local mean free path is much 
larger than the size of the considered domain, they will stream out of the system, practically without 
further interactions.

The global behaviour of the radiation we are interested in requires that, when computing the optical depth
according to Eq.~\eqref{eq:general optical depth}, a) we can 
start the path from any point inside the domain, 
and b) the final point can be \textit{any} point on the boundary of 
the computational domain (or possibly in a region transparent to radiation at that wavelength, 
where boundary conditions also apply).
Among all the possible paths connecting $A$ to the boundary, the statistical interpretation
suggests to consider the paths that tend to minimize the optical depth (i.e., the paths with the least
number of interactions) as the most likely ways for radiation to escape. 

Following these prescriptions, we define the optical depth of a point $\v{x}$ as
\be\label{eq:tau x}
\tau(\v{x})=\min_{\{\gamma|\gamma:\v{x}\rightarrow \v{x}_{\rm e}\}}
\int_\gamma \frac{\d s}{\lambda(s)},
\ee 
where $\v{x}_{\rm e}$ is one point from which radiation can escape freely. 
The calculation of the optical depth, therefore, implies finding a point
$\v{x}_{\rm e}$ and a path $\gamma$ that together tend to minimize the value of $\tau$ 
as given by Eq.~\eqref{eq:general optical depth}.
In what follows, we present our prescription for finding $\gamma$ and $\v{x}_{\rm e}$ in MODA.
Strictly speaking, the solution provided by the algorithm is not necessarily the one 
that minimizes Eq.~\eqref{eq:general optical depth}. However, the search for a global minimum is
the guiding principle in constructing, step by step, the path $\gamma$ towards the boundary of the
domain\footnote{We 
note that a slight overestimation of the minimum optical 
depth is not necessarily incorrect, even considering the assumptions of our method. 
The (theoretical) minimum is the smallest value that
$\tau$ could \textit{possibly} take (anything below that is nonphysical), but (slightly) larger 
values are of course possible and perhaps to be expected in nature.}. 

The main algorithm, presented in the next section, is based on the 
following assumptions:

\bi
\i[(a)] $\lambda$ is the only quantity that determines 
$\tau$, see Eq.~\eqref{eq:tau x};
the input for our algorithm is therefore a certain spatial
distribution $\lambda(\v{x})$, which can be
either analytically known or obtained from a simulation;
\i [(b)] $\lambda$ is a smooth function of position;
\i [(c)] the boundaries of the computational domain are 
completely transparent regions, and on large scales $\lambda$
increases with the distance from the center of the computational domain.
However, there can be local variations and local maxima and minima.
\ei

\noindent Note that depending on the context one may wish to either 
compute a spectral or an average (gray) mean free path. Moreover, 
depending on the radiation--matter interaction processes involved, 
$\lambda$ may be either a scattering or an effective total mean free path 
\citep[see, for example, ][Eq.~14.5.57]{raffelt2001,shapiro83}.

\subsection{Algorithm}
\label{sec:method algorithm}

\begin{figure*}[ht!]
\centering
\includegraphics[width=\textwidth]{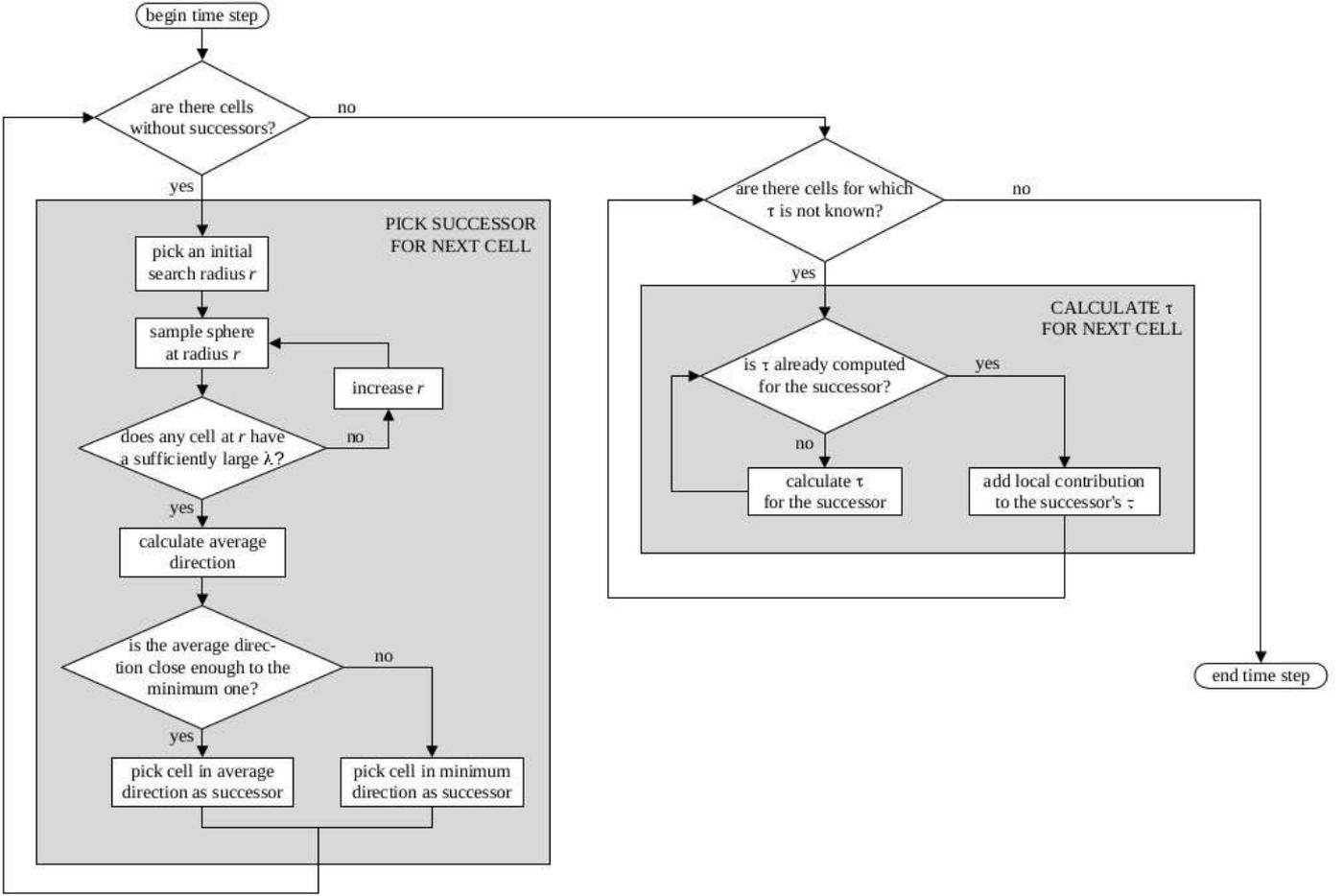}
\caption[Flow chart of the MODA algorithm]
{\textbf{Flow chart of the MODA algorithm} -- This schematic representation of MODA
reveals two distinct steps (the choice of successors and the integration of $\tau$), each of which
involves a loop over all grid points (or cells). The algorithm is completely described by this chart,
and all specific decisions that are not included in the figure (e.g., how to pick the initial $r$ and how
much to successively increase it afterwards) are parametrizations that depend on the underlying implementation 
(i.e., grid-based or tree-based).}
\label{fig:flow chart}
\end{figure*}

From the form of Eq.~\eqref{eq:tau x} we see that the path that minimizes the optical
depth favours regions where the mean free path $\lambda$ is larger.
If $\lambda$ were a monotonic function of the radial position, 
ever increasing towards the edges of the computational domain, the algorithm would be trivial: 
each cell would ``pick'' its neighbour with the largest $\lambda$ as the direction of integration,
which would guarantee that the optimal solution is achieved.

In practice, however, $\lambda$ is not always monotonic: non-trivial astrophysical
scenarios often contain discontinuities, shocks, clumps of hot dense matter,
all of which yield a highly anisotropic and non-homogeneous profile of $\lambda$.
In such situations, always choosing the local maximum of $\lambda$ leads 
to getting ``trapped'' in regions that \textit{are} local maxima.
On a larger scale, however, the assumption that $\lambda$ increases towards the
edge of the computational domain is still expected to hold. The MODA algorithm makes 
use of this assumption by searching for \textit{sufficiently larger} mean free paths, first locally,
and if such $\lambda$'s are not found then at increasingly larger radii. By parametrizing
what ``sufficiently larger'' means, one is able to obtain a balance between accuracy
and speed.

For convenience, since Eq.~\eqref{eq:tau x} contains $\lambda$ in the denominator,
we introduce the \textit{opacity} $\invlam$
\be\label{eq:inverse lambda}
\invlam(\mathbf{x})= {1}/{\lambda(\mathbf{x})},
\ee 
which is the quantity we use in the code instead of $\lambda$
(since $\lambda\rightarrow\infty$ while
$\invlam\rightarrow 0$ towards the edges of the computational domain).

To illustrate the algorithm, let us assume we are interested in the optical depth at a point $\v{x}$.
We first define a sphere of radius $r_1$ centered around $\v{x}$, and on its surface $\mathcal{S}(\v{x},r_1)$ we search 
for a point $\v{z}$ that satisfies
\be\label{eq:MODA condition f}
\invlam(\mathbf{z}) < f_{\rm dec} \invlam(\mathbf{x}),
\ee 
where $f_{\rm dec} \lesssim 1$ is a parameter\footnote{Typical 
values of $f_{\rm dec}$ that we have tested (see \sec{sec:tests} 
for more details) lie in the interval $0.1 - 0.8$. 
Larger values, closer to unity, can be less effective in avoiding 
local extrema, while smaller values require more computational effort, 
and are prone to push the decreasing length scale search too early towards the 
edge of the computational domain, introducing boundary conditions effects. 
In any case, we recommend a careful examination of $\invlam(\mathbf{x})$ and 
its typical spatial and temporal variations, in order to choose 
an adequate value for $f_{\rm dec}$.}.
If such a point is found then the vector $\hat{\v{e}}_{\v{v}}=\v v/|\v v|$, with $\v v \equiv \v{z}-\v{x}$,
gives the direction of integration. Otherwise, we keep extending our search to 
spheres of increasing radii $r_i$, with $i=2,3,...$ until a point $\v{z}$ that satisfies
Eq.~\eqref{eq:MODA condition f} is found. The choice of transparent
boundary conditions ensures that this always happens. Once the direction of integration 
$\v v$ is found, we simply select --from all the direct neighbours of $\v{x}$-- the cell
$\v x'$ that is in the direction of $\v v$, which we call the ``successor'' of $\v{x}$.
Once $\v{x}'$ is known, the search for its own successor can be performed and so on
until  a point $\v{x}_{\rm e}$ on the boundary is reached.
The sequence of $n$ intermediate points connecting $\v{x}$ and
$\v{x}_{\rm e}$, \{$\v{x}^{(0)} \equiv \v{x}$, $\v{x}^{(1)} \equiv \v{x}'^{(0)}$, \dots, $\v{x}^{(i)} \equiv 
\v{x}'^{(i-1)}$, \dots, $\v{x}^{(n+1)} \equiv \v{x}_{\rm e}$ \}, provides the needed path, $\gamma$.
Along this path, the optical depth $\tau(\mathbf{x})$ is calculated as:
\be\label{eqn: optical depth in MODA}
\tau(\mathbf{x})
= \sum_{i=1,n+1} \, 
\left(  \int_{\gamma^{(i)}: \mathbf{x}^{(i-1)} \rightarrow \mathbf{x}^{(i)}} \; \invlam \, {\rm d}s \right).
\ee

%
%
\section{Implementation}
\label{sec:implem}

In the following, we describe a possible implementation of the MODA algorithm
on a static grid and in a tree-structure. 
The discussion is accompanied by a flow chart that summarizes all the steps
of the algorithm (\figref{fig:flow chart}). In order to keep this description as general
as possible, we collect the more technical details in \seca{sec:app technical}.

In the previous section, we presented the basic concepts focusing on a single starting
point. A straightforward implementation would be to successively apply that procedure to each point
inside the computational domain. However, this approach is not computationally efficient, since it
does not employ a relevant aspect of the problem: given two points, there is a non negligible 
probability that, depending on the spatial distribution of $\invlam$, their two paths towards the
edge of the domain merge at a certain
point. Given the unambiguous character of the prescription to find the successor, the optical depth
of the common part will be the same, and it will be equal to the optical depth of the 
intersection point. In order to take advantage of this, we divide the implementation
of the algorithm into two distinct steps (apart from preliminary operations, such as setting boundary conditions), 
each applied to the whole domain:
1) finding the relevant distances for the decrease of 
$\invlam$, the integration directions, and the successors; 2) performing the integration.
Keeping these steps separate ensures that common paths are not computed multiple times.

\subsection{Implementation into a grid code (``grid-MODA'')}
\label{sec:implem grid}

As an example, we implement the algorithm into the Eulerian, uniform mesh MHD code
FISH \citep{kaeppeli11}. The implementation into different types of grid codes (non-uniform 
and/or non-Cartesian) is straightforward and requires only a moderate amount of changes.

\subsubsection{Setting boundary conditions}
\label{sec: setting boundary conditions}
Setting boundary conditions implies, first of all, providing a predefined value for $\tau$ at the edges of
the grid: 
\be
\label{eq:tau boundary}
\tau(\v{x}_{\rm e}) = \tau_{\rm boundary}.
\ee
Safe values for the boundary condition are
$\tau_{\rm boundary} \lesssim h / \lambda_{\rm boundary}$,
where $h$ is the local resolution length (cell size) and $\lambda_{\rm boundary}$ is the 
typical value of the mean free path at the boundary.
Since the method requires exploration of neighbouring areas, special attention must
be given to cells located close to the boundary, for which the spherical surface $\mathcal{S}$ may extend
beyond the edges of the domain.
There are several possible solutions to this problem;
one would be to modify Eq.~\eqref{eq:MODA condition f} 
to search for a direction $\v z$ that satisfies the opposite condition,
\be\label{eq:MODA condition f boundary}
\invlam(\mathbf{z}) > f_{\rm dec}^{-1} \invlam(\mathbf{x}),
\ee 
and then to choose the direction $\v{e}_{\v{v}}$ by setting $\v{v} = - (\v{z} - \v{x})$; 
Another solution would be to set a boundary layer thick enough to prevent one from
accessing cells that do not exist.

\subsubsection{Finding distances, directions and successors}
\label{sec: find distances, directions and successors}
We begin searching for points that satisfy Eq.~\eqref{eq:MODA condition f} at a distance
$r_1\sim h$, i.e. comparable to the local resolution length $h$.
Instead of calculating the intersection of the grid with the spherical surface $\mathcal{S}(\v{x},r_i)$ 
of center $\v{x}$ and radius $r_i$
(which would become increasingly expensive for high $i$'s), we sample the surface in a predefined
set of $m$ ``equidistant'' points, $ S_{m}(\v{x},r_i) = \{ \v{z}_j = \v{x} + r_i \v{y}_j \}_{j=1,m} $,
where $\left| \v{y}_j \right| \simeq  1$.
Once a point $\v{z}_{\rm min}$ that satisfies Eq.~\eqref{eq:MODA condition f}
is found inside $S_m(\v{x},r_i)$, the radius $r_i$ becomes the length scale over which 
$\invlam$ decreases down to the chosen limit $f_{\rm dec}\invlam(\mathbf{x})$. 

One could already select $\v v_{\rm min}=\v{z}_{\rm min}-\v{x}$ as the direction of integration,
but this was found to sometimes lead to abrupt 
discontinuities in the optical depth.
To avoid this effect, it is possible to devise a smoothed direction of integration 
$\v v_{\rm avg}$, by using all
sampling points inside $S_m(\v{x},r_i)$ (or even a subset) and weighting each respective
unit direction vector, $\v{y}_j$, with a suitable monotonically decreasing function of $\invlam$, 
$w(\invlam)$:
\be\label{eq:moda smoothed direction}
\mathbf{v}_{\rm avg}(\mathbf{x}) = \left( 
\sum_{S(\v{x},r_i)} \, \v{y}_j \, w(\invlam(\v{y}_j)) \right) 
 / \left| \sum_{S(\v{x},r_i)} \, \v{y}_j \, w(\invlam(\v{y}_j)) \right| .
\ee
This average favors the directions where $\invlam$ is minimum on the 
spherical surface, but it also takes into account the behaviour 
of $\invlam$ at the scale of $r_i$, smoothing the 
evolution of the direction in case of sharp, local variations. 
In order to avoid compensation effects due to the averaging itself
(which would for instance be the case along the axis of symmetry in highly symmetrical configurations, 
where the average may cancel out the contributions from symmetrical points 
with the same $\invlam_{\rm dec}(\mathbf{x})$),
we calculate the cosine
of the angle between $\v{v}_{\rm avg}$ and $\v{v}_{\rm min}$.
If the two directions are too different 
(i.e., if the cosine is below a threshold, 
for example $\cos \theta_{\rm lim}=0.9$)
we reject $\mathbf{v}_{\rm avg}(\mathbf{x})$, calculated with 
Eq.~\eqref{eq:moda smoothed direction}, 
and use $\mathbf{v}_{\rm min}(\mathbf{x})$ instead.
The definition of $\cos \theta_{\rm lim}$ sets the maximum allowed 
discrepancy between the two directions.

Once the direction of integration $\v{v}$ (given by either $\v{v}_{\rm avg}$ or $\v{v}_{\rm min}$)
has been found, we select the closest cell to $\v{x}$ in the direction given by $\v{v}$, from
an appropriate set of neighbours, and denote it by $\v{x}'$.
It is noteworthy that, although the search 
for the integration direction $\v{v}_{\rm dec}(\v{x})$ 
is made by ``looking ahead'' to whatever distance is necessary in 
order to find an acceptable $\invlam$, the integration itself is done in small increments, comparable
to the local resolution. The choice of the appropriate set of neighbours, among which $\v{x}'$
has to be searched, has to balance two opposing tendencies: including cells more distant
from $\v{x}$ increases the angular resolution in searching for $\v{x}'$, but at the same time
decreases the accuracy in the path discretization.

In a grid based code determining $\mathbf{x}'$ is relatively simple and efficient, 
since grid cells have fixed indices based on their geometrical position.
We note that all operations involving a cell refer
in fact to the geometrical center of the cell. This introduces discretization
errors, since most of the times 
the condition $\v{z}-\v{x}=\v{v}$ cannot be fulfilled exactly
(the only exception being paths which traverse entire rows or columns of cells, 
or which go exactly along diagonals).
The impact of this discretization on the resulting path will be shown and explained later on,
in \figref{fig: RCB-MODA-example}.

Another caveat is that closed circular paths may occur, even though ``looking ahead'' for the 
direction decreases the chances
of this to happen. Nevertheless, whenever a successor is chosen for a given cell one
\textit{must} check whether it leads to a loop in Eq.~\eqref{eqn: optical depth in MODA}.
If that is the case the next best cell must be picked as a successor.
In the tests we have performed, closed loops occurred rarely and usually in presence of
very symmetric local extrema of $\invlam$. Thus the error introduced by not choosing 
exactly the direction that tends to minimize the optical depth is usually small and infrequent.
The inspection of the integration paths, necessary to discover the presence of loops, 
has a minor effect on the performance of the algorithm, especially compared with 
the more demanding searches of the distances and of the directions.

\subsubsection{Performing the integration}
\label{sec: integration}
Once the direction of integration has been found throughout the entire computational domain,
each cell falls into one of these two categories: either it is a boundary point,
in which case $\tau$ is already known there, or it is a point inside the computational domain,
in which case one knows in what direction radiation moves away from it. 
The actual integration takes the form of a loop through all the cells: those for which $\tau$ is known
are simply skipped, while for the others the integration path is reconstructed
by following the list of successors, until arriving to a cell with known $\tau$.
The integral in Eq.~\eqref{eqn: optical depth in MODA} then becomes a sum of partial $\Delta \tau$'s,
\be
\tau(\v{x}) \approx \tau(\v{x}^{(l)}) + \sum_{i=1}^{l} \invlam_{(i,i-1)} \, |\v{x}^{(i-1)}-\v{x}^{(i)}| 
\label{eq:moda tau real integration},
\ee 
where $\invlam_{(i,i-1)}$ is an average between $\invlam(\v{x}^{(i)})$ and 
$\invlam(\v{x}^{(i-1)})$, $\v{x}^{(0)} = \v{x}$, and $\tau(\v{x}^{(l)})$ is the first already-computed 
value of the optical depth along the path $\gamma$ (or, in case $\gamma$ does not overlap with 
an already-computed path, $\v{x}^{(l)} = \v{x}_{\rm e}$).

\subsection{{Implementation using a tree-structure (``tree-MODA'')}}
\label{sec:implem SPH}
Our goal is to implement the MODA algorithm into
an SPH code (for recent reviews of this method see e.g. \citealt{monaghan05} 
and \citealt{rosswog09}). Most SPH codes make use of hierarchical structures (trees) 
to search for particle neighbours 
and calculate gravitational forces. In what follows, we will describe the implementation of MODA into a
recently developed recursive coordinate bisection (RCB) tree
\citep{gafton11}. The implementation strategy, however, is not restricted to this
type of tree and can straightforwardly be adapted to other tree types such
as octree \citep{barnes86} or other binary trees \citep{benz90}.

The RCB tree is not built down to the last particle, but down to small groups of adjacent particles
that are always aggregated into what we call ``lowest-level cells'' (or ll-cells),
also found in the ``tree literature'' as ``leaves'' \citep{oxley03,springel05,gaburov10,hubber11,clark12} 
or as ``buckets'' \citep{dikaiakos96,stadel01,wadsley04}. 
Since the average number
of particles per ll-cell (typically $\sim 12$) is much smaller than the 
average neighbour number ($\sim 100$), the size
of any ll-cell will always be smaller than the smoothing length
of its particles. Since the smoothing length is the typical
length scale over which physical quantities are resolved in SPH, 
one does not expect the mean free path to exhibit large variations within any one
ll-cell. It is therefore unnecessary to calculate the optical depth 
$\tau$ for each individual particle, and one expects the same results by
computing it \textit{per ll-cell}. This idea has also been followed by \citet{oxley03},
whose radiative transfer scheme for SPH uses tree leaves as the radiating and absorbing elements,
and by \citet{stamatellos05}, who use cells whose linear size is comparable to the local smoothing length.

In principle, the algorithm for a tree follows the steps presented in
\sec{sec:implem grid}: successively larger radii are used for searching for a sufficiently
larger mean free path, and once the correct direction is found, the successor of 
the current cell is chosen as the closest cell in the respective direction.
In our tree-adapted version of the MODA algorithm, ``grid cell'' translates into ``ll-cell''. 
Since cells located at a certain radius cannot be identified by simply operating
on cell indices, as in the case of grid-based codes, 
we rely on the tree walk infrastructure of the RCB tree
\citep[{\S}2.2]{gafton11} to return a list of ll-cells located 
within a certain spherical shell of radius \textit{r} and parametrized width \textit{dr}.
Furthermore, once the direction of integration is known, one cannot use a predetermined
list of directions, described by known angles, to quickly find the closest cell
in the necessary direction. One must instead compute the angle between the directions
to the surrounding cells (which can in principle be pre-computed upon building the tree) 
and the desired direction of integration, and then pick the best (i.e., the largest cosine). 
On the other hand, once the successors are known for all ll-cells, the integration step is virtually
identical to that performed by grid-MODA.

%
%
\section{Tests, applications and performance}
\label{sec:tests}

In this section, we apply both versions of MODA to two types of three dimensional configurations: 
first, to cases with analytically known inputs and, when possible, solutions (cases 1-3),
and, second, to practical astrophysical applications (cases A-D).
The former (\sec{sec:tests analytic}) are validation tests for some simple 
cases where $\lambda$ and $\invlam$ are analytic functions of the Cartesian coordinates,
while the latter (\sec{sec:tests astroph}) are based on snapshots 
from real astrophysical simulations.
The solutions compared in each case were obtained with some of the following methods:
(a) analytic solution, if available;
(b) grid-MODA;
(c) tree-MODA;
(d) local RbR method; 
(e) global RbR method.
In \tabref{table: tests} we summarize the cases we have explored and the available solutions.

\begin{table}[ht!]
\centering
   \begin{tabular}{|l||c c c|c c c c|}
     \hline
    Input $\rightarrow$  & \multicolumn{3}{|c|}{Analytic $\invlam$} & \multicolumn{4}{|c|}{Astrophysical $\invlam$}  \\ \hline
    $\downarrow$ Method  & 1  & 2  & 3  & A  & B  & C & D  \\ \hline
    analytic $\tau$      & x  & x  &    &    &    &   &    \\ \hline
    grid-MODA            & x  & x  & x  & x  & x  &   &    \\ \hline
    tree-MODA            & x  & x  & x  &    &    & x & x  \\ \hline
    local RbR            & x  & x  & x  &    & x  &   &    \\ \hline
    global RbR           & x  & x  & x  & x  &    &   &    \\ \hline
   \end{tabular}
\caption{Summary of all the configurations we have analyzed and the available solutions for each of them.
See the text for more details. In tests 1 and 2 the analytic and the global ray-by-ray solutions coincide.}
\label{table: tests}
\end{table}

In the local RbR method, we integrate $\invlam$ along a set of predefined straight paths 
(rays) from each point of the grid, and choose the 
optical depth at that point to be the minimum amongst these integrals. To obtain a local
directional resolution comparable with the one provided by MODA, we choose 98 directions, passing by
the centers of the neighbouring cells satisfying Eq.~\eqref{eqn: neighbouring cells}.
In the global ray-by-ray method, we interpolate $\invlam$ from the Cartesian mesh to a spherical one, 
calculate the local $\tau$ by integrating $\invlam$ along the radial path, 
and interpolate $\tau$ back on the Cartesian mesh. The spherical mesh is characterized by
30 polar and 60 azimuth angles.

\subsection{Analytic tests}
\label{sec:tests analytic}
The simplest possible tests for MODA involve analytic,
spherically symmetric distributions $\invlam(r)$, 
see Eq.~\eqref{eq:inverse lambda}.
For such tests, the direction of integration is known 
(it can only be the outward radial direction, by symmetry),
and the optical depth $\tau$ can therefore be computed analytically. These are
also the most insightful tests since the error can be computed for
each individual cell or particle, and the reason for each deviation
can usually be pin-pointed and understood, allowing us to acknowledge the inherent limitations of MODA.
Tests 1 and 2 below are based on two such configurations, while test 3 involves a more
complex (non-spherically symmetric, though still analytic) spatial configuration.

For grid-MODA, the computational domain is an equally-spaced Cartesian
grid of $600^3$ cells, with each cell having unitary width.

For tree-MODA, we represent the $\invlam$ profile by $\sim 10^5$ (in tests 1 and 2)
or $2\times 10^6$ (in test 3) particles.
As a conservative precaution we use $\sim 1$ particle per ll-cell.
Where this is not possible due to the tessellation algorithm, we prescribe the $\lambda$ of an 
ll-cell to be the average of the $\lambda$'s of all its particles.\\

\begin{figure}[ht!]
\centering
\includegraphics[width=\linewidth]{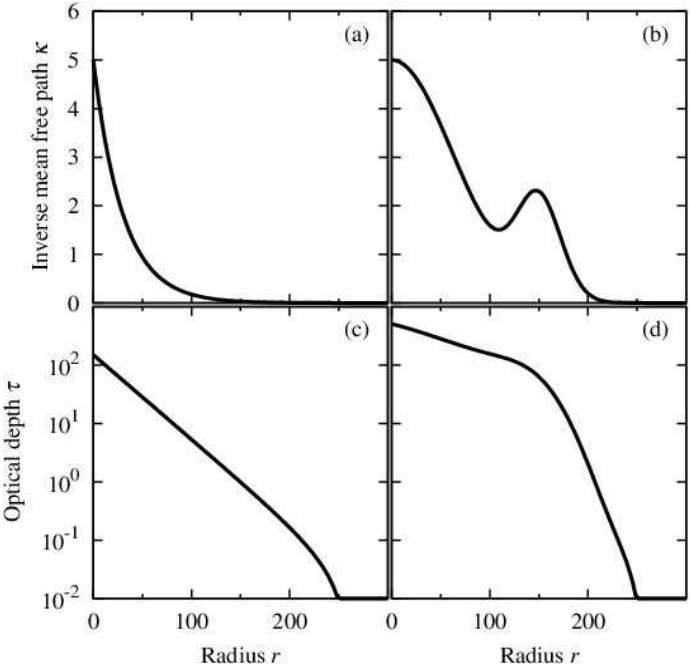}
\caption{\textbf{Analytic $\invlam$ and $\tau$ for tests 1 and 2} --
This plot shows the spherically symmetric initial conditions $\invlam(r)$ 
and optimal results $\tau(r)$ for the first two analytic tests: 
(a) monotonic $\invlam(r)$ as given by Eq.~\eqref{eq:sigma analytic 1}; 
(b) non-monotonic $\invlam(r)$ as given by Eq.~\eqref{eq:sigma analytic 2};
(c) $\tau(r)$ as given by Eq.~\eqref{eq:tau analytic 1};
(d) $\tau(r)$ as given by Eq.~\eqref{eq:tau analytic 2}.
}\label{fig:analytic-plots}
\end{figure}

\textit{Test 1: monotonic, spherically symmetric}.

For the first and simplest test 
we define a monotonically decreasing spherically symmetric inverse mean free path
\be\label{eq:sigma analytic 1}
\invlam(r)=\invlam_0 \exp\left(-\frac{r}{d_0}\right),
\ee 
where $r$ is the radial distance from the center of the
computational domain, which is $(0,0,0)$ in our Cartesian coordinate system. 
For our tests we use $\invlam_0=5$, $d_0=30$, which leads
to the distribution shown in \figref{fig:analytic-plots}(a),
with the solution being the analytic function
\be\label{eq:tau analytic 1}
\tau(r)=\begin{cases}
\invlam_0 \; d_0 \left(e^{-r/d_0}-e^{-\rout/d_0}\right)+\tau_{\rm boundary} &, r<\rout \\
\tau_{\rm boundary} &, r\geq\rout
\end{cases}
\ee 
as shown in \figref{fig:analytic-plots}(c).
The boundary conditions are defined by the cutoff radius $\rout=249$ and 
the optical depth of the transparent regime, $\tau(r>\rout)=10^{-2}$.
We use a parameter $f_{\rm dec}=0.8$, see Eq.~\eqref{eq:MODA condition f}.

\begin{figure*}[ht!]
\centering
\includegraphics[width=10cm]{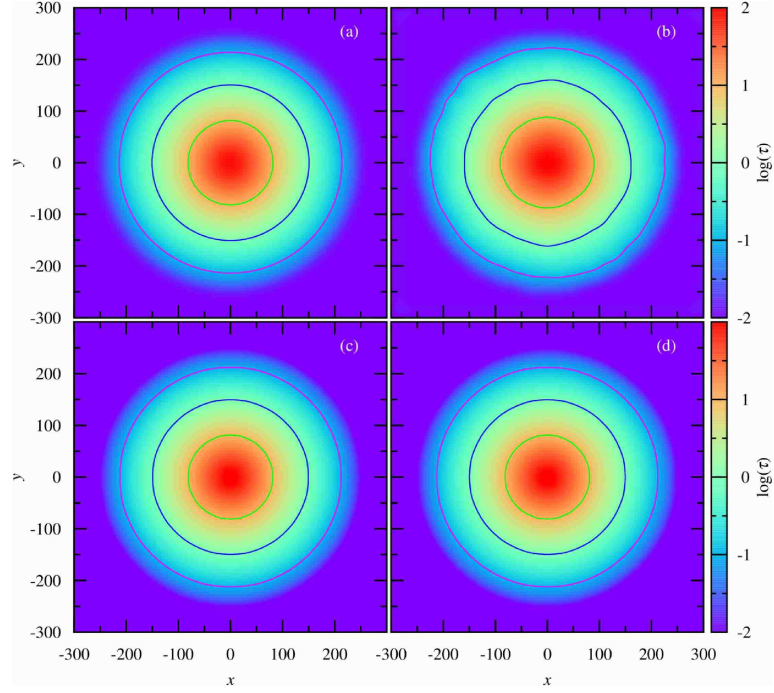}
\caption[Results of test 1]
{\textbf{Results of test 1} -- Logarithm of optical depth $\tau$ for
the first (monotonic, spherically symmetric) analytic test, as obtained with four methods: 
(a) analytic solution [see Eq.~\eqref{eq:tau analytic 1} and \figref{fig:analytic-plots}(c)];
(b) tree-MODA;
(c) grid-MODA;
(d) local ray-by-ray.
Contour lines correspond to $\log\tau=-1,0,1$. Qualitatively, the plots are in agreement. 
The small deviations from spherical symmetry in the contours on the tree-MODA plot
are largely due to interpolation to the grid used for plotting.
}\label{fig:tau-result analytic 1}
\end{figure*}

In \figref{fig:tau-result analytic 1} we present the resulting optical depth calculated with
different methods. The fact that contours of $\tau$ are concentric circles shows that all
algorithms accurately reproduce the original spherical symmetry of the input data.
While spherical symmetry is directly encoded in the analytic solution and in the global ray-by-ray method,  
\figref{fig:tau-result analytic 1}(a), it is less obvious for MODA, where no symmetry and 
no predefined path is assumed.

The larger error of tree-MODA (especially near the boundary $\rout$,
where $\tau$ is very low) is both due to the lower resolution compared to the grid code ($600^3$ vs $10^5$)
and due to the interpolation on the regular grid used for plotting.
In astrophysical simulations, the former issue can be considerably 
attenuated by sampling the fluid well enough, i.e. by making sure
that all particles have a sufficiently large number of neighbours and that
the smoothing lengths are well below the relevant physical scales.\\

\textit{Test 2: non-monotonic, spherically symmetric}.

The second test is also spherically symmetric, but has a non-monotonic
feature (a local maximum) halfway through the computational domain,
which simulates non-monotonic profiles that often appear in real simulations,
as discussed in \sec{sec:method algorithm} and as will be clearly seen in 
the astrophysical tests (\sec{sec:tests astroph}).
This distribution is defined as the superposition of
a Gaussian centered in the origin and an off-centered Gaussian,
\be\label{eq:sigma analytic 2}
\invlam(r)=\invlam_0 \exp\left(-\frac{r^2}{d_0}\right)+\invlam_1 \exp\left(-\frac{(r-r_1)^2}{d_1}\right).
\ee 
For our tests we use the parameters 
$\invlam_0=5$, $\invlam_1$=2, $d_0=8000$, $d_1=1000$, $r_1=150$ in order
to get the distribution shown in \figref{fig:analytic-plots}(b).
The integration of Eq.~\eqref{eq:sigma analytic 2} 
can be performed analytically, with the result being
\be
\tau(r)=\begin{cases}
\begin{array}{ll}
\frac{\sqrt{\pi}}{2}&\left\{
\sqrt{d_0}\invlam_0
\left[-\erf\left(\frac{r}{\sqrt{d_0}}\right)+\erf\left(\frac{\rout}{\sqrt{d_0}}\right)\right)\right.\\
&\phantom{\left\{\right.}+
\sqrt{d_1}\invlam_1
\left.\left(-\erf\left(\frac{r-r_1}{\sqrt{d_1}}\right)+\erf\left(\frac{\rout-r_1}{\sqrt{d_1}}\right)\right]\right\}
\\
& \; + \; \tau_{\rm boundary},\quad r<\rout \end{array}&\\
\tau_{\rm boundary},\quad r\geq\rout&
\end{cases}
\label{eq:tau analytic 2}
\ee 
where $\erf(x)$ is the Gauss error function. This leads to the profile $\tau(r)$
shown in \figref{fig:analytic-plots}(d). 
The same boundary conditions apply as before, i.e. $\rout=249$, $\tau(r>\rout)=10^{-2}$, 
and $f_{\rm dec}=0.8$.

\begin{figure*}[ht!]
\centering
\includegraphics[width=10cm]{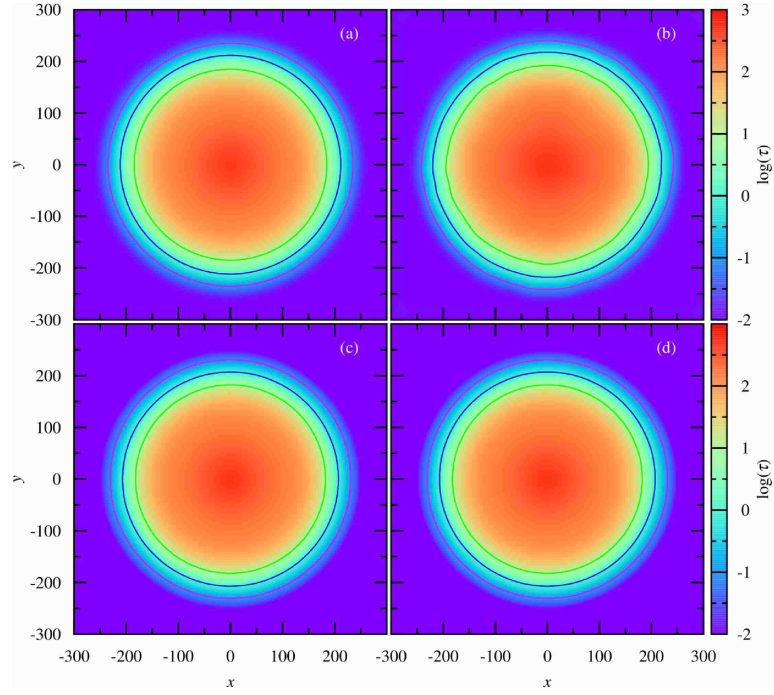}
\caption[Results of test 2]
{\textbf{Results of test 2} -- Logarithm of optical depth $\tau$ for
the second (non-monotonic, spherically symmetric) analytic test, as obtained with four methods: 
(a) analytic solution [see Eq.~\eqref{eq:tau analytic 2} and \figref{fig:analytic-plots}(d)];
(b) tree-MODA;
(c) grid-MODA;
(d) local ray-by-ray.
Contour lines correspond to $\log\tau=-1,0,1$.
}\label{fig:tau-result analytic 2}
\end{figure*}

The results are presented in \figref{fig:tau-result analytic 2} and are consistent with 
those in the previous section, with all three methods yielding excellent results, even though
tree-MODA is slightly worse near the boundaries, where the SPH resolution is poorest.\\

\textit{Test 3: asymmetric.}

For the third model we define a superposition of two off-centered, elliptical
Gaussian distributions
\bea
\invlam(x_1,x_2,x_3)&=&\sum_{n=1,2}\invlam_n \exp\left[
-\left(\frac{x_1-\tilde x_{1,n}}{s_{1,n}}\right)^2\right.\nonumber\\
&&\left.-\left(\frac{x_2-\tilde x_{2,n}}{s_{2,n}}\right)^2
-\left(\frac{x_3-\tilde x_{3,n}}{s_{3,n}}\right)^2
\right],\label{eq:sigma analytic 3}
\eea 
with $\invlam_1=120$, $\invlam_2=80$, $s_{1,1}=35$, $s_{2,1}=30$, $s_{3,1}=25$,
$s_{1,2}=45$, $s_{2,2}=15$, $s_{3,2}=30$, $\tilde{x}_{1,1}=\tilde{x}_{2,1}=\tilde{x}_{3,1}=40$, and
$\tilde{x}_{1,2}=\tilde{x}_{2,2}=\tilde{x}_{3,2}=-40$. 
The same boundary conditions apply, i.e. $\rout=249$ and $\tau(r>\rout)=10^{-2}$. This time we use $f_{\rm dec}=0.5$,
since the more complicated geometry requires more accuracy and larger decreases in the distance determination.

\begin{figure*}[ht!]
\centering
\includegraphics[width=10cm]{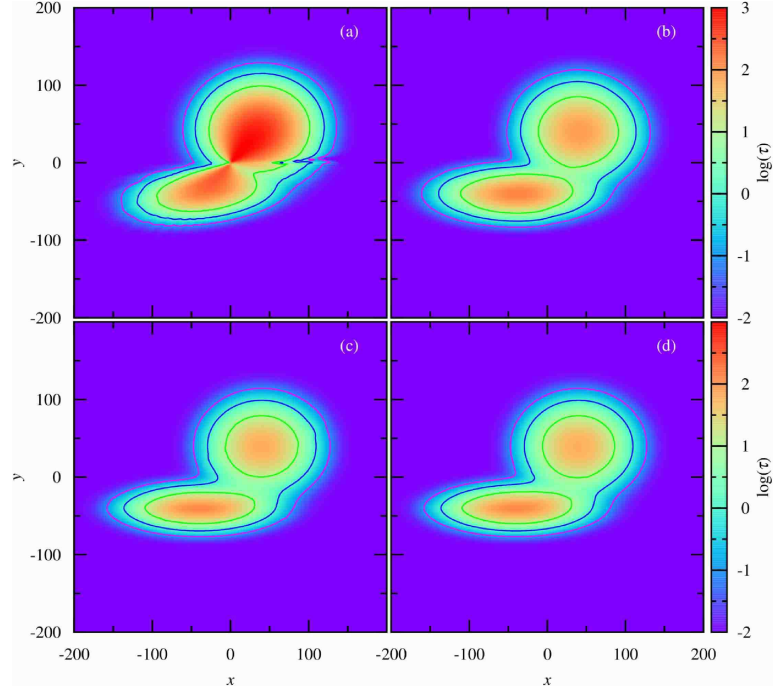}
\caption[Results of test 3 ($z=0$ cross-section)]
{\textbf{Results of test 3 ($z=0$ cross-section)} -- Logarithm of optical depth $\tau$ for
the third (non-spherically symmetric) analytic test, as obtained with four methods: 
(a) global ray-by-ray;
(b) tree-MODA;
(c) grid-MODA;
(d) local ray-by-ray. The plots represent a cut through the $z=0$ plane.
Contour lines correspond to $\log\tau=-1,0,1$.
}\label{fig:tau-result analytic 3 xy}
\end{figure*}

\begin{figure*}[ht!]
\centering
\includegraphics[width=10cm]{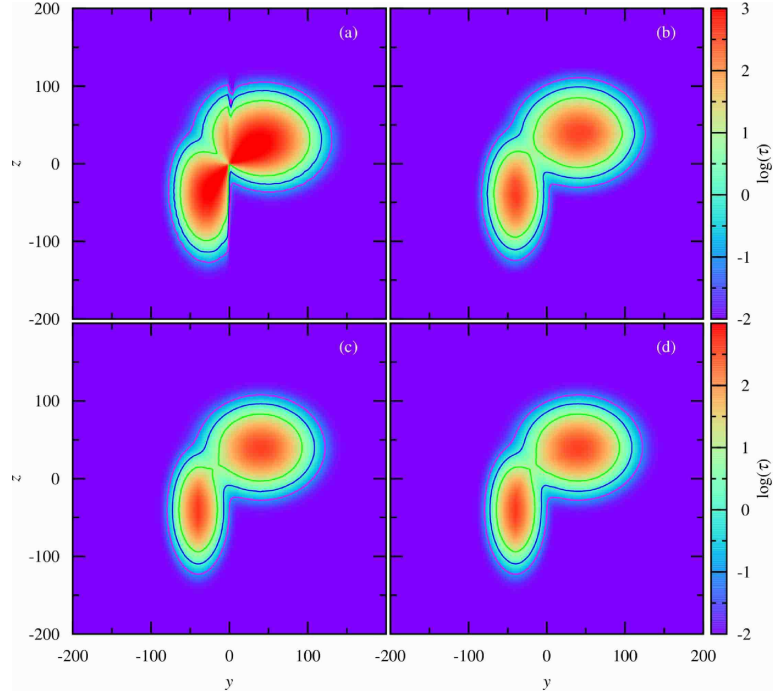}
\caption[Results of test 3 ($x=0$ cross-section)]
{\textbf{Results of test 3 ($x=0$ cross-section)} -- Logarithm of optical depth $\tau$ for
the third (non-monotonic, spherically symmetric) analytic test, as obtained with four methods: 
(a) global ray-by-ray;
(b) tree-MODA;
(c) grid-MODA;
(d) local ray-by-ray. The plots represent a cut through the $x=0$ plane.
Contour lines correspond to $\log\tau=-1,0,1$.
}\label{fig:tau-result analytic 3 yz}
\end{figure*}

Two sets of results are shown, corresponding to cross-sections of the $z=0$ plane 
(\figref{fig:tau-result analytic 3 xy}) and of the $x=0$ plane
(\figref{fig:tau-result analytic 3 yz}).
The integration of Eq.~\eqref{eq:sigma analytic 3} is not straightforward to perform
analytically, since the direction of integration is not radial, but highly dependent 
on the position. For this reason, we use the results obtained with the local RbR 
method as reference, \figref{fig:tau-result analytic 3 xy}(d) and 
\figref{fig:tau-result analytic 3 yz}(d). In contrast to the previous cases,
the global ray-by-ray method [\figref{fig:tau-result analytic 3 xy}(a) and 
\figref{fig:tau-result analytic 3 yz}(a)],
shows inaccuracies and artifacts due to the lack of spherical symmetry of $\invlam$.
On the other hand, the other tests are qualitatively consistent and show remarkable agreement and 
similar resolution.

Two important considerations, common to all the tests we have performed, have to be pointed out. 
First, the optical depth is a physical
quantity that can usually span a wide range of values (five orders of magnitude in this example); therefore,
large relative errors in a limited number of zones (particularly near the edges, where $\tau$ is essentially zero) 
do not imply a noticeable effect on the overall calculation of $\tau$. 
Second, MODA provides a comparable
result at a considerably lower computational cost than the local RbR method 
(see \sec{sec:performance}).

\subsection{Astrophysical applications}
\label{sec:tests astroph}
We now turn to real astrophysical scenarios, and we choose
snapshots of multidimensional simulations 
with highly asymmetrical and non-isotropic configurations,
that span orders of magnitude in density.
 
In order to apply our algorithm, we consider thermodynamical conditions extracted 
from 3D simulations and, based on them, we calculate the physical $\lambda$
for electron neutrinos with a specific energy (chosen to be 54 MeV).
Once the $\lambda$ and hence the $\invlam$ distribution is obtained, 
all subsequent steps are identical to those from the analytic tests.

\subsubsection{Grid-MODA}
\label{sec:tests astroph grid}

\textit{Case A: core-collapse supernova}\label{sec:tests astro A}
\begin{figure*}[ht!]
\centering
\includegraphics[width=12cm]{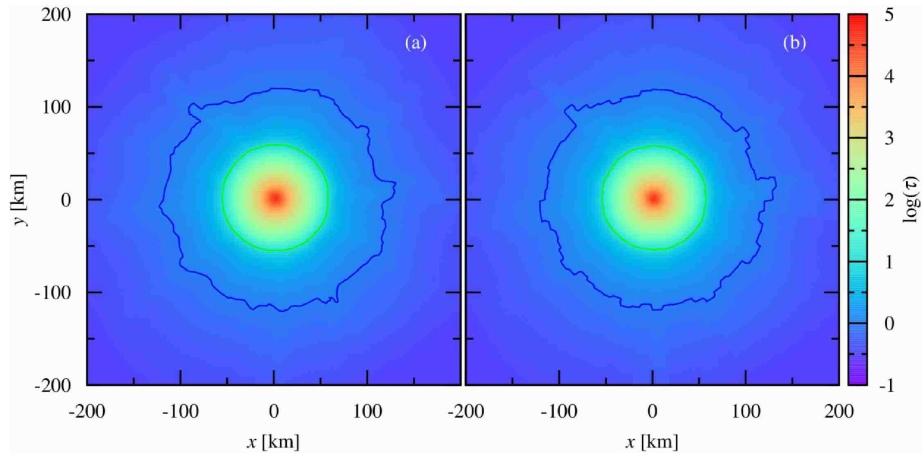}
\caption[Results of case A]
{\textbf{Results of case A} -- Logarithm of optical depth $\tau$ for
the astrophysical case A, as obtained with two methods: 
(a) grid-MODA;
(b) global ray-by-ray.
Contour lines correspond to $\log\tau=-1,0,1$.
}\label{fig:tau-result astroph A}
\end{figure*}

In the first application, we take a 2D Cartesian slice of data (density, temperature and 
electron fraction), from a 3D core collapse supernova simulation performed with the 
ELEPHANT code \citep{whitehouse08}, with a spatial resolution of 2 km. 
Considering that the system is, to a first approximation, spherically symmetric, we 
calculate the electron neutrino optical depth using grid-MODA and the global 
ray-by-ray method (with the polar angle discritized by 64 angular bins). Both results are shown in 
\figref{fig:tau-result astroph A}. We find similar results with both methods: inside a radius of 
about 60 km (which includes the proto-neutron star and the inner part of the shocked material), 
the optical depth contours are spherically symmetric, as expected. 
Above that radius, matter is convective and  multi-dimensional effects come into play. 
The contour lines start to show multi-dimensional features, which are related to 
non-homogeneous features in density, temperature and electron fraction. The main difference 
between both methods is the spatial resolution of the results. While the ray-by-ray 
approach decreases its resolution moving outwards, the new algorithm maintains a 
constant resolution. As a result, the outer optical depth contours are smoother 
than those found by the global ray-by-ray method.\\

\textit{Case B: neutron star merger remnant}\label{sec:tests astro B}
\begin{figure*}[ht!]
\centering
\includegraphics[width=15.5cm]{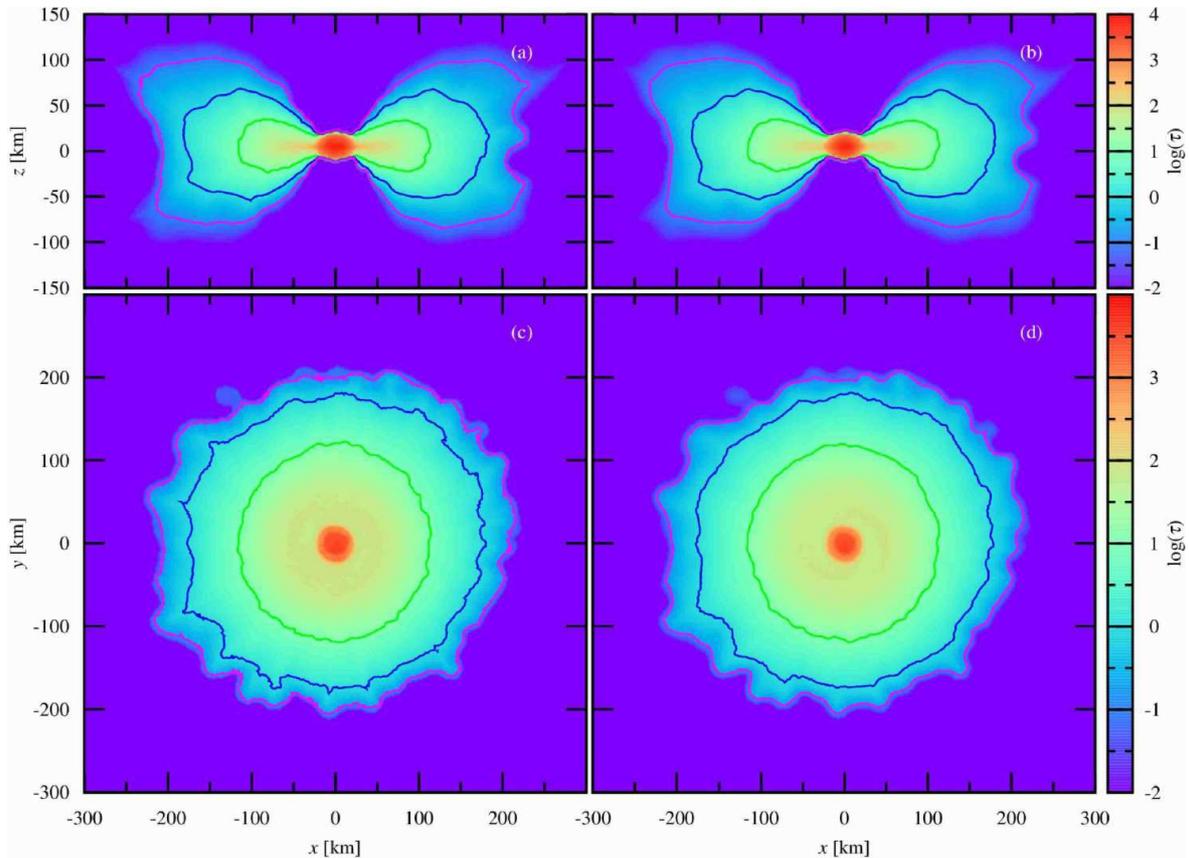}
\caption[Results of case B ($z=0$ cross-section)]
{\textbf{Results of case B} -- Logarithm of optical depth $\tau$ for
the astrophysical case B, as obtained with two methods: 
(a)\&(c) grid-MODA;
(b)\&(d) local ray-by-ray.
Top (bottom) panels refer to $y=0$ ($z=0$) cross-sections.
Contour lines correspond to $\log\tau=-1,0,1$.
}\label{fig:tau-result astroph B}
\end{figure*}

As a second application case we take the matter distribution (density, temperature and electron fraction) 
from the remnant of a double neutron star merger simulation \citep{price06}, 
performed with the SPH code MAGMA \citep{rosswog07}, and map the particles onto a 3D Cartesian, 
equidistant grid with spatial resolution of 2 km.
The calculation of the optical depth is performed with grid-MODA and 
the results are shown in \figref{fig:tau-result astroph B}(a)\&(c), 
along with the results of a local RbR method (b)\&(d). 
The two plots are in good agreement, though fine-structure differences do appear.
We suspect that these differences are
mainly due to the fact that our method is based on a purely local exploration of the suitable 
radiation path, and does not invoke any special global symmetry or predefined direction. 
In this sense, our 3D method provides an optical depth calculation that is more 
general and automatically adapts to the geometry of the matter distribution. 
The high resolution, three dimensional local RbR method is, also in this case, 
prohibitively expensive from 
a computational point of view, and the resolution shown in this test is never 
achieved with this method in actual hydrodynamic calculations.

\subsubsection{Tree-MODA}
\label{sec:tests astroph SPH}

The input for the tree-MODA tests was produced with a version of the MAGMA code \citep{rosswog07}.
From the densities and electron fractions of the SPH particles we computed the mean free paths 
for $E=54$~MeV neutrinos.

We present the results of tree-MODA alone for two reasons. First,
we do not have another particle-based code that computes optical depths 
for SPH simulations. Second, due to the intrinsic resolution adaptivity 
that characterizes SPH codes, it would not be straightforward to compare those results
with the ones obtained by grid-MODA, after having remapped the SPH matter distribution 
on a uniform grid.\\

\textit{Case C: white dwarf collision}\label{sec:tests astro C}
\begin{figure*}[ht!]
\centering
\includegraphics[width=11cm]{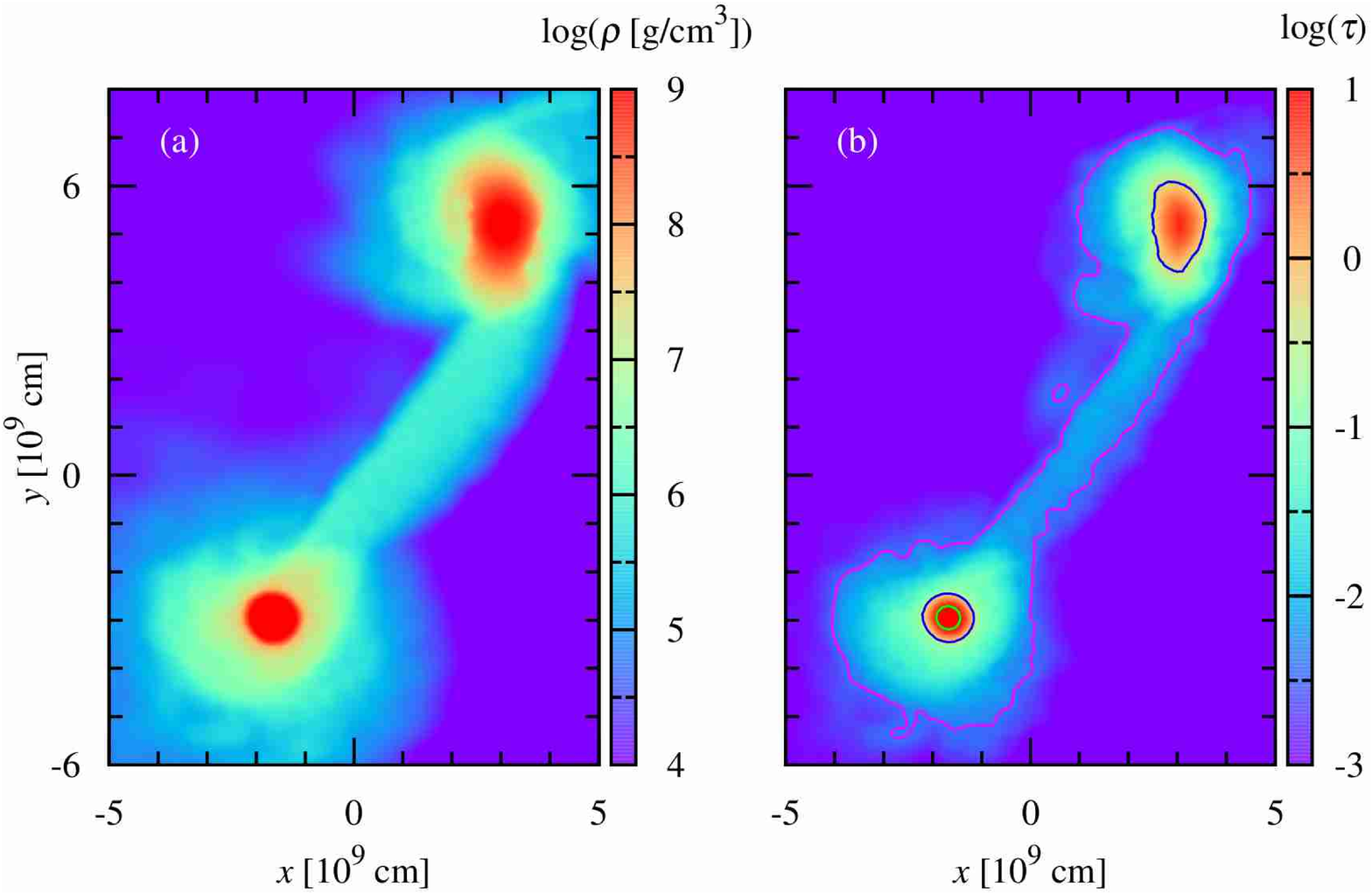}
\caption[Results of case C]
{\textbf{Results of case C} -- (a) Density $\rho$ and (b) logarithm of optical depth $\tau$ 
for the astrophysical application C. The density profile is taken directly from the simulation output,
while $\tau$ is computed with tree-MODA. The SPH resolution is $N\sim 6\times 10^5$ particles.
The plot shows a $z=0$ cross section. 
Contour lines correspond to $\log\tau=-2.4,-0.3,1$.
}\label{fig:tau-result astroph C}
\end{figure*}

The first astrophysical application of tree-MODA uses a snapshot from an
off-center collision between a 0.6 M$_{\odot}$ and a 0.9 M$_{\odot}$ white dwarf
\citep{rosswog09a} from a simulation with  $\sim 6\times 10^5$ SPH particles. 
The density profile and the resulting optical depth are shown in 
\figref{fig:tau-result astroph C} for a $z=0$ cross section.
At this stage, the stars had a first close encounter and are now moving 
towards their apocentre separation before they fall back towards each other again.
Some debris shed in the collision is aligned in the form of a temporary 
``matter bridge'' between the two stars. Note that the secondary (near $y=6 \times 10^9$ cm) 
is heavily spun up by tidal torques.
Since the $E=54$~MeV neutrino mean free path is comparable to the size of the white dwarfs,
one could expect the maximum optical depth (at the center of the star) to be around 1.
The results of our calculations are in agreement with that: the contour lines of $\tau$ for the
heavier star are concentric circles and have a maximum value of $\sim 1$;
the lighter star has a much smaller optical depth, and the tidal bridge is essentially transparent
and comparable in $\tau$ to the  debris halos surrounding the two stars.\\

\textit{Case D: neutron star collision}\label{sec:tests astro D}
\begin{figure*}[ht!]
\centering
\includegraphics[width=16cm]{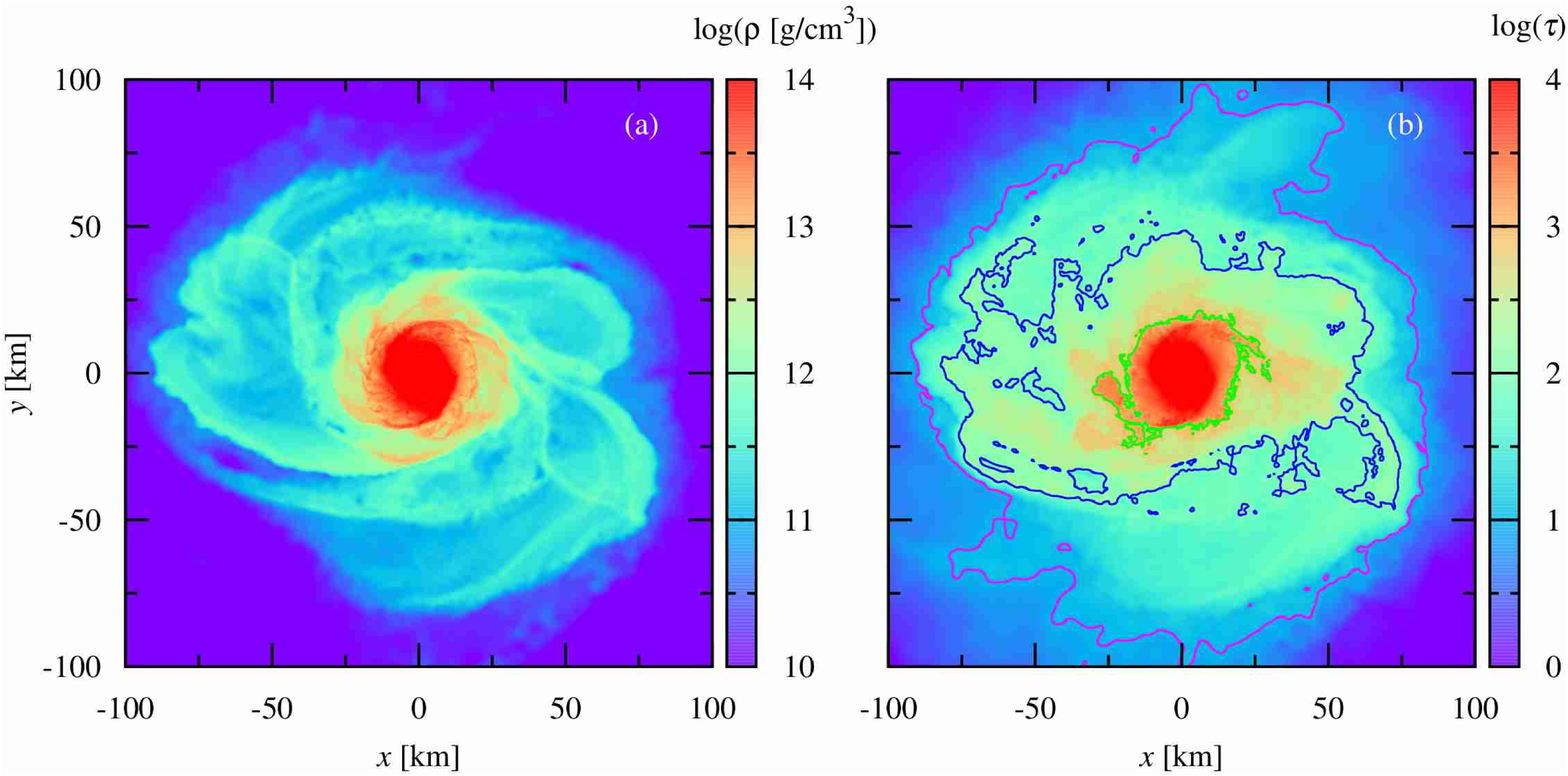}
\caption[Results of case D]
{\textbf{Results of case D} -- (a) Density $\rho$ and (b) logarithm of optical depth $\tau$
for the astrophysical application D. The density profile is taken directly from the simulation output,
while $\tau$ is computed with tree-MODA. The SPH resolution is $N\sim 8\times 10^6$ particles.
The plot shows a $z=0$ cross-section, and only the inner $200$~km of the disc ($\gtrsim 1000$~km)
are shown.
Contour lines correspond to $\log\tau=1,2,3$.
}\label{fig:tau-result astroph D}
\end{figure*}

The second application for tree-MODA is a snapshot from a high resolution ($\sim 8 \times 10^6$ particles)
simulation of a collision of two neutron stars of 
masses 1.4 M$_{\odot}$ and 1.3 M$_{\odot}$ \citep{rosswog13}.
The density profile and the resulting optical depth are shown in 
\figref{fig:tau-result astroph D} for a $z=0$ cross section.
The snapshot provides a late-stage picture of the encounter, when, after several close passages
a pulsing and rapidly rotating central remnant, surrounded by shocked debris from several mass shedding
episodes has formed.
The entire structure has a radius of $\gtrsim 500$~km, but we show just the inner $100$~km,
where most of the mass is concentrated and the debris contains a number of  fine structures from 
the mass shedding and subsequent shocks. 
This is a highly non-trivial, non-isotropic and asymmetric problem. Still, 
the resulting optical depth profile found by MODA accurately portrays the complex 
debris structure.

\subsection{Performance}
\label{sec:performance}

\begin{figure*}[ht!]
\centering
\includegraphics[width=12cm]{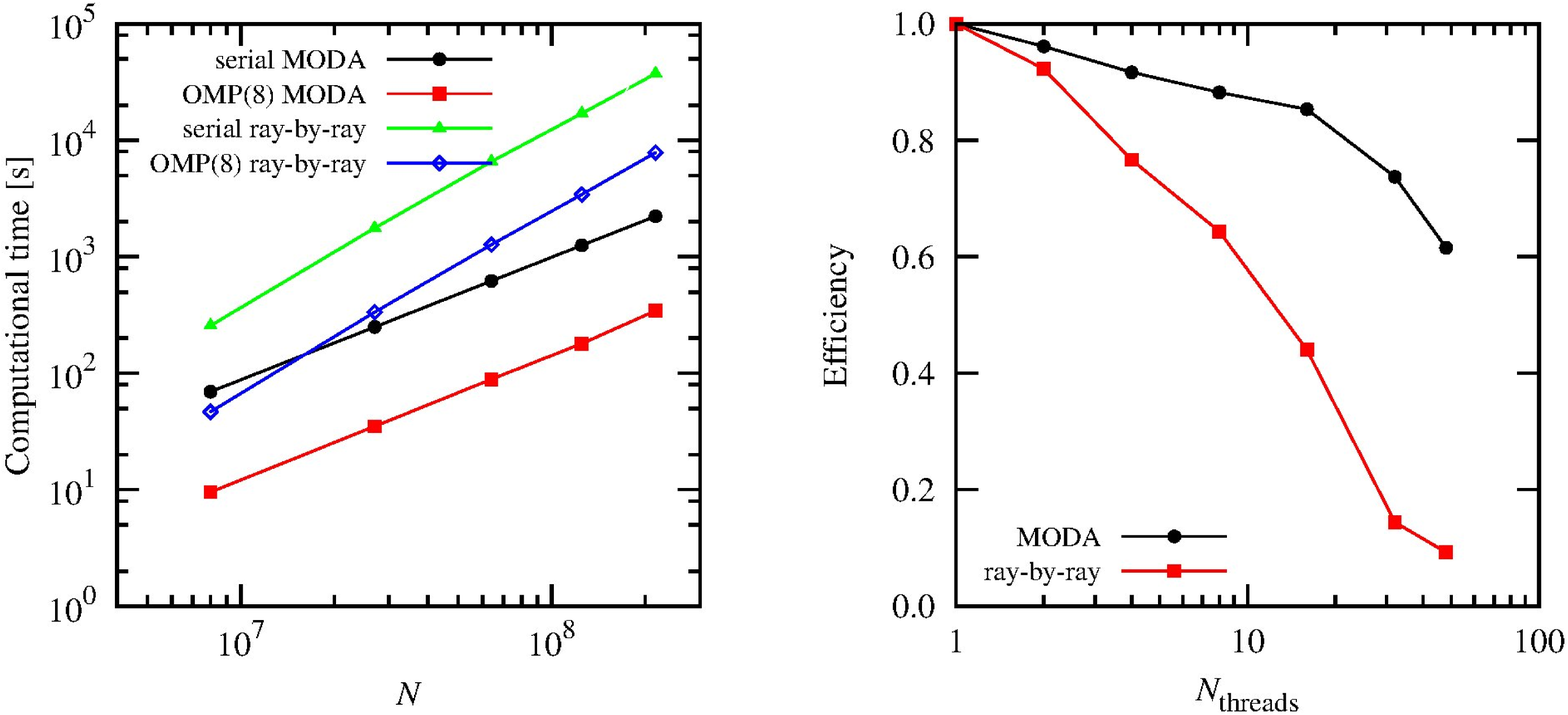}
\caption[Performance comparison between MODA and the ray-by-ray algorithm]
{\textbf{Performance comparison between MODA and the ray-by-ray algorithm} -- 
(left) Computational time as a function of the effective number of cells $N$, obtained with 
both the serial codes and their OpenMP versions using 8 threads. (right) Computational efficiency
as a function of the number of OpenMP threads. For these
parallelization tests, the dimension of the computational domain is fixed to $N=400^3$.
}\label{fig:scaling}
\end{figure*}

In this section, we compare the performance of grid-MODA and of the local
ray-by-ray method, the two grid-based methods that are able to handle complex matter distributions
without specific symmetries. We focus on three-dimensional simulations, where the
computational effort is higher and, consequently, the impact of the optical depth
calculation cost is larger. In order to compare the scalings with the grid size
we employ a computational domain of size $N=N^{1/3}\times N^{1/3}\times N^{1/3}$
and compare both the serial and the OpenMP versions with $N_{\rm threads}=8$.
As test case we use analytic test 1 (see \sec{sec:tests analytic}).
The values of $N^{1/3}$ range from $200$ to $600$ for all cases and
the results are presented in \figref{fig:scaling}(a).
In addition of being a more general method that automatically adapts to the geometry
of the matter distribution, MODA is usually more than one order of magnitude faster than the local
ray-by-ray method (in the serial as well as the parallelised versions), and exhibits
better scaling with increasing $N$ ($t_{\rm MODA} \sim N^{1.75}$ while $t_{\rm RbR} \sim N^{2.31}$).

In the case of MODA, we notice that most of the computational time ($\sim 70 \%$) is
spent to find, for each point of the domain, the relevant distance
where $\invlam$ decreases enough to satisfy Eq.~\eqref{eq:MODA condition f}.

Most loops appearing in MODA have been parallelized from the beginning using OpenMP constructs.
Its scaling with the number of OpenMP threads, shown in \figref{fig:scaling}(b) by the so-called efficiency
$\eta = T_1 / \left( p T_p \right)$ (where $T_1$ is the serial computational time and
$T_p$ is the computational time using $p$ threads)
is much better than the efficiency of a ray-by-ray algorithm.

%
%
\section{Conclusions}
\label{sec:discussion}

In this paper we presented a new, efficient algorithm for the calculation of
optical depths $\tau$ from any given profile of the mean free path $\lambda$. 
There are no restrictions related to the symmetry or the configuration of the 
computational domain. All that is needed is that $\lambda$ be globally increasing
towards the edges of the domain, which is normally a fair assumption.
The algorithm was implemented both in a grid-based and a tree-based code,
and proved to be suitable for both Eulerian and Lagrangian methods.
The three-dimensional tests we presented, starting with analytic, spherically-symmetric
configurations and ending with highly anisotropic and heavily shocked configurations,
proved that the algorithm provides excellent accuracy in multi-dimensional calculations, 
while being less computationally expensive than more traditional ray-by-ray methods.

%
%
\section*{Acknowledgements}
AP was supported by the Helmholtz-University Investigator grant No. VH-NG-825; he
thanks also the Stockholm University for its hospitality in June 2013, 
and the University of Basel for its hospitality in January 2014, 
during the final stages of the development of MODA. 
EG thanks the University of Basel for its hospitality during the summer of
2011, when the work on the SPH version of MODA was initiated. 
EG and SR acknowledge the use of computational resources provided by 
the H\"{o}chstleistungsrechenzentrum Nord (HLRN), and the use of SPLASH 
developed by D. Price to visualize SPH simulations.
RC and ML acknowledge the support from the HP2C Supernova project and the ERC grant FISH.
AP, RC and ML thank the use of computational resources provided by 
the Swiss SuperComputing Center (CSCS), under the allocation grants s412 and s414.

%
%
\bibliography{MODA}

\appendix

%
%
\section{Technical implementation details}
\label{sec:app technical}

In this appendix, we provide more technical details about our actual implementation
of MODA in a static Cartesian grid (\sec{sec:app technical grid})
and in a tree-code (\sec{sec:app technical SPH}).

\subsection{Grid code}
\label{sec:app technical grid}

\textit{Setting boundary conditions}.
Concerning the setting of the boundary conditions, in our grid implementation of MODA
the definition of a thick boundary layer has been chosen, since it is easier
to implement and faster to execute. However it has the small disadvantage of reducing
the size of the useful computational domain: in the boundary layer, the optical depth
takes by definition a constant, predefined value $\tau_{\rm boundary}$, according to 
Eq.~\eqref{eq:tau boundary}.
This may not be a problem since the choice of a constant is based on the 
(generally valid) assumption that the area near the edge of the computational domain
is completely transparent to radiation (i.e., $\lambda_{\rm boundary} \gg h$).
The width of the layer is assumed to be 8\% of the linear size of the computational box.
The standard value used in MODA for the boundary optical depth is $\tau_{\rm boundary}=10^{-2}$.

\textit{Finding the distances, the directions and the successors}.
The investigation of the inverse mean free path on a sphere of radius $r_i$ and center $\v{x}$
is the first step in the search for $\v{x}'$, the local successor point of $\v{x}$. 
As described in \sec{sec: find distances, directions and successors}, this is done
by sampling the spherical surface with a set of $m$ equidistant points, 
$ S_{m}(\v{x},r_i) = \{ \v{x} + r_i \v{y}_j \}_{j=1,m} $,
where $\left| \v{y}_i \right| \simeq  1$.
While in two dimensions these would simply be $m$ points 
on a circle separated by angles of $2\pi/m$ radians; in three dimensions, we choose $\{ \v{y}_j \}_{j=1,m}$ 
as the solutions of the Thomson problem \citep{thomson04,wales2006} on a unitary sphere.
In all our tests, we adopted $m=64$.

The function $w(\invlam)$, used in Eq.~\eqref{eq:moda smoothed direction} to devise a 
smoothed direction of integration is $w(\invlam) = 1/\invlam$.
The maximum allowed angular distance between $\v{v}_{\rm min}$ and $\v{v}_{\rm avg}$ is set by
$\cos \theta_{\rm lim} = 0.9$.
After the determination of the ideal integration direction $\v{v}$, the successor point has 
to be found out of a suitable set of neighbours.
For this set, we consider all cells 
 $\v{z}(i',j',k')$ around $\v{x}(i,j,k)$ that satisfy
\be
 \label{eqn: neighbouring cells}
|i-i'|\leq 2,\qquad
|j-j'|\leq 2,\qquad
|k-k'|\leq 2
\ee 
(here $i$, $j$ and $k$, eventually primed, are integers representing the Cartesian coordinate 
of the cell within the grid). In the cases where two cells have their centers aligned with $\v{x}$, we choose
the one closest to $\v{x}$. In total, the number of neighbours is 98.

Since angular operations are easier 
in spherical coordinates, one begins by extracting the 
direction $(\theta,\varphi)$ of the vector $\v{v}$.
Then one loops through the neighbouring cells. 
In this loop, one discards all the cells that are too ``far'' from the
direction $\varphi$, which is not expensive to check: since one already knows the relative
position of the surrounding cells, one can easily store and retrieve 
the pre-computed spherical components of $\v{z}-\v{x}$. 
Once the list of cells is culled,
the spherical distance between 
all normalized direction vectors 
$(\v{z}-\v{x})/|\v{z}-\v{x}|$ and the normalized direction $\v{\hat v}$ is computed, 
and the closest
cell in the direction of $\v{v}$ is chosen. 

\textit{Performing the integration}.
In the calculation of Eq.~\eqref{eq:moda tau real integration}, $\lambda_{(i,i-1)}$ is computed 
as the arithmetic average between $\invlam(\v{x}^{(i)})$ and $\invlam(\v{x}^{(i-1)})$. 
For a uniform grid, this corresponds to the application of the trapezoidal rule for the calculation of $\Delta \tau$.

\subsection{SPH code}
\label{sec:app technical SPH}

\begin{figure}[ht!]
\centering
\includegraphics[width=.4\textwidth]{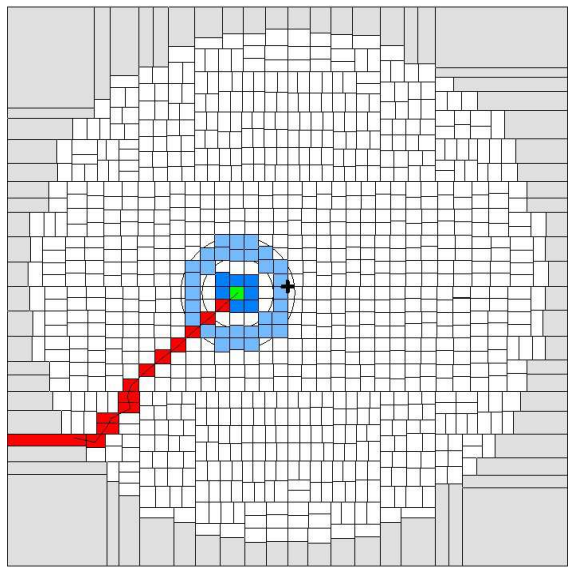}
\caption[Example of a path found by tree-MODA]
{\textbf{Example of a path found by tree-MODA} -- 
This simple test for a two-dimensional Sobol distribution \citep[see][\ssymb{7.7}]{press92} with
$10^4$ particles (1024 ll-cells) and a spherically symmetric, monotonic mean free path
profile (see \sec{sec:tests analytic}) reveals most of the 
inner workings and the problems of the algorithm. See the text in \seca{sec:app technical SPH} for details.}
\label{fig: RCB-MODA-example}
\end{figure}

We illustrate the steps of the tree-MODA algorithm in \figref{fig: RCB-MODA-example},
which uses a very basic, two-dimensional, spherically symmetric computational domain with
very low resolution, and presents a typical path found by tree-MODA. All steps
discussed below can be traced on the figure.

\textit{Setting the boundary condition.}
The boundary condition for the optical depth, described by Eq.~\eqref{eq:tau boundary},
is assigned in a similar way as for the grid-based code, although tree-MODA does not
need such a broad boundary (since only cells that are literally touching the edges of
the computational domain, i.e., the gray cells in \figref{fig: RCB-MODA-example}, need to be assigned the 
boundary conditions).

\textit{Empty ll-cells.}
Due to the tessellation provided by the tree code, empty cells may appear anywhere 
in the computational domain. For spatially-balanced trees, which create cells according to
a prescription independent of the particle distribution, 
it is obvious why empty cells appear \citep[see Figure 1 in][]{barnes86}.
In density-balanced trees, on the other hand, they appear in low-density regions 
and are compensated for by ``over-populated'' ll-cells in dense regions, 
since the RCB tree enforces by construction an ``average'' number of particles per ll-cell. 
In general, the lower the ``average'' number of particles per ll-cell and
the less homogeneous the system is, the higher the chances are of empty ll-cells being produced.
As the RCB tree is built proactively, all empty cells 
are effectively compressed into zero volume and then ignored in all tree walks and during the 
integration of $\tau$, because they do not contain useful information (i.e., particles).

\textit{Finding the correct direction of integration}.
A tree walk --similar to the neighbour search tree walk-- performed for each ll-cell $A$
(in the figure, a fiducial cell shown in green) returns a list of ll-cells that are 
closer to $A$ than a certain value $dx$ (the dark blue cells). 
By setting \textit{dx} to an infinitesimally small
value we obtain the simplest possible test, in which the cells \textit{must} touch
(\textit{dx} should not be exactly zero due to the imprecise nature of floating point
arithmetic, see e.g. \citealt[{\S}4.2 of the second volume]{knuth73}, or \citealt{goldberg91}).
In practice we allow \textit{dx} to be a little larger, 
a fraction of the typical size of the cell, since the tessellation may, in rare cases, 
produce cells which are extremely close to each other but do not quite touch.

Once the list of neighbouring cells is obtained, one loops through them and checks whether 
Eq.~\eqref{eq:MODA condition f} holds for any of them. 
If it does, the one with the largest mean free path is picked as successor. 
Otherwise, the search radius is extended and another tree walk is performed, which
returns a list of ll-cells (in the figure, the light blue cells) 
whose centers of mass fall within a spherical shell of radius $r_2$ and width $dr$ 
(marked by the two black circles).
The operation continues until a suitable cell is found. Since all boundary cells count as
``suitable'', this will always happen eventually.

We note that the search radius and its increments are not fixed throughout the
computational domain, like in grid-MODA,
but are functions of the local smoothing length. This is necessary because of 
the adaptive resolution of SPH: in dense regions smoothing
lengths are small and we only need to slightly extend the search radius in order to
sample more ll-cells -- and implicitly particles, while in less dense regions smoothing
lengths are large and we need to look further away in order to sample the same number
of cells.

The algorithm returns a (correct) straight path for as long as the cells are 
aligned just like in a grid, but it cannot do so once the underlying cell structure becomes irregular. 
When that happens, the path deviates from the expected radial path and the (minimum) optical
depth is over-estimated.

A more subtle error of the algorithm (which can be only alleviated by increasing the resolution) 
can be observed as follows. 
If one draws the radial line from the center of the computational domain (black cross) 
to our fiducial green cell,
one sees that the path found by tree-MODA is not in fact along the radial direction (which in this
case would be close to horizontal). This is a discretization problem caused by
the low resolution of the mesh of ll-cells. One notices that there are eight direct neighbours
at $r_1$. Because the profile of $\lambda$ is radial, MODA will choose the lower-left corner cell (since
its center of mass is further away than that of e.g. the cell directly to the left, 
it also has larger $\lambda$), even though this may not lead to
the correct analytic solution. The same problem appears at radius $r_2$: because of the discretization,
the spherical shell is not uniformly filled with cells, and there are cells whose 
center of mass ``just doesn't make it'' into the shell 
(e.g. immediately to the right of the domain center). MODA \textit{will} select the best possible
solution from the list of cells and most of the times it will provide a very reasonable approximation,
but one has to keep this in mind as being an important source of error.

\textit{Finding the closest cell in the correct direction}.
If a cell with a large enough mean free path has been found amongst the very first neighbours
(this can happen if the value of $f$ in Eq.~\eqref{eq:MODA condition f}
is very small, which in practice is a bad idea since it may lead to getting trapped
in regions of local maxima, as explained in 
\sec{sec:method algorithm}) then that cell is chosen as a successor.

Otherwise, it means that we have a direction of integration (given by the coordinates of the cell found
in the previous step) and need to find the closest ll-cell in the respective direction.
Achieving directed transport has long been a problem in SPH: 
radiation propagates along straight lines,
while a typical SPH particle distribution is highly irregular:
if one wants to find the closest cell in a given direction, the least complicated
approach is to compute the direction vector to each neighbouring cell, and then
the cosine of the angle between that vector and the desired direction of integration.
The cell which gives the largest cosine will be the best approximation 
to the desired direction. 

If one implements tree-MODA at the level of particles instead of ll-cells, 
an interesting alternative is the concept of virtual SPH particles \citep{pawlik08}:
instead of searching for an actual particle that is only approximately
in the desired direction of integration, 
a virtual particle is created \textit{exactly} in the direction where it is needed, 
at a distance comparable with the local resolution length, allowing the path to 
proceed in \textit{exactly} the desired direction.
Implementing this into the current code is not straightforward: the path cannot stop at the newly-introduced 
virtual particle, so after its creation one needs to find its successor as well; 
this means that one would need to maintain two separate lists, 
a list of ll-cells and a list of virtual particles, etc.

\textit{Performing the integration}.
Once the direction of integration has been found throughout the entire computational domain
the integration is performed in a very similar manner to the grid-MODA, by means
of a recursive subroutine.
For each ll-cell, the successor is known and can fall in one
of two categories: either its optical depth $\tau$ is not yet known,
in which case it is recursively computed, or it is known, in which case
Eq.~\eqref{eq:moda tau real integration} is evaluated with $l=1$, i.e. by just computing
the increase in the optical depth when moving between the two cells,
and then adding it to the already-computed optical depth of the successor.

%
%
\section{Parallelization}
\label{sec:app parallelization}

\textit{Parallelization of grid-MODA.} 
Most loops appearing in MODA have been parallelized from the beginning using OpenMP constructs.
However, the integration of MODA into a complex MPI code such as FISH \citep{kaeppeli11} 
requires special attention.
By far the most challenging issue is that calculating the optical depth $\tau$
(and here we include both the determination of an integration path, 
and the evaluation of the integral itself),
requires information from the entire computational domain. 
This is problematic in an MPI environment, where the information about the computational
domain is distributed between multiple nodes, each with its own separate memory.
The repeated exchange of necessary information between all nodes would be highly expensive,
and a MODA-like algorithm that only needs to exchange information about the boundary zones
between neighbouring sub-domains has not yet been devised.
On the other hand, the calculation of $\tau$ \textit{only} depends on the mean free path,
which is a local quantity that can be computed by each MPI node separately; 
moreover, the calculation of $\tau$ is efficient and has been OpenMP parallelized.
The natural solution is then to set aside one node 
(i.e. the ``MODA node'') for the sole purpose of the optical depth
calculation. This node receives a list of mean free paths from all processors, calculates
the optical depth using the MODA algorithm, and then sends back the list of optical depths to
each respective processor.

Initial experiments have shown that the calculation of the optical depth for one energy bin
in core collapse supernova simulations with FISH takes of the order of one MHD time step.
Since $\sim 20$ energy bins are typically used, and both the effective and the total optical
depth (see \sec{sec:method definitions assumptions}) need to be calculated, 
a few possibilities exist to balance the time spent in these two parts of the code:\\
(a) The optical depth can be assumed to not change significantly after one MHD time step,
therefore $\tau$ can be evaluated once every few time steps; for example, 
if computing $\tau$ for each of the 20 energy bins takes $\sim$ one MHD time step, then
the optical depth will effectively be calculated once every 20 time steps,
which depending on the simulation may be sufficient.\\
(b) Optical depths are dependent on the neutrino energy, however we can assume that their paths
do not change significantly between two neighbouring energy bins.
It may therefore suffice to only calculate the integration path (which is the most
expensive part of MODA) for three representative energy values 
(``low'', ``medium'', ``high''), and
then for each of the 20 energy bins to choose which of the three paths is more appropriate
and perform the integration along it -- using the correct 
spectral mean free paths.\\
(c) In addition, one can also use multiple MODA nodes, e.g. three nodes, one for each
energy regime described in (b), or two nodes, one for the effective and one for the 
total optical depth, etc.

\textit{Parallelization of tree-MODA.} 
The four parts of the tree-MODA algorithm described in 
\seca{sec:app technical SPH} are --at this stage-- parallelized with OpenMP.
The tree walks that find the successor of each cell greatly resemble those 
performed by the SPH neighbour search and the gravity calculations,
in that they are performed independently for each ll-cell. This makes the algorithm
ideal for parallelization with OpenMP.
The integration of the optical depth is also easy to parallelize, although it
may happen that, if several threads work at the same time on cells
that have partially overlapping integration paths, the calculation for the
common part will be done multiple times.
\end{document}